\documentclass[secnumarabic,amssymb, amsmath,12pt, nofootinbib,tightenlines, nobibnotes, aps, prl]{revtex4}
\usepackage{graphicx}
\usepackage{cancel}
\usepackage{xcolor}
\begin{document}
\newcommand{\be}{\begin{equation}}
\newcommand{\ee}{\end{equation}}
\newcommand{\ba}{\begin{eqnarray}}
\newcommand{\ea}{\end{eqnarray}}
\newcommand{\bea}{\begin{eqnarray*}}
\newcommand{\eea}{\end{eqnarray*}}
\newcommand{\nn}{\nonumber}
\newcommand{\mpi}{m_{\pi}}
\newcommand{\mta}{m_{\tau}}
\newcommand{\tpp}{$\tau^- \to \pi^-\pi^0\nu_{\tau}$}
\newcommand{\eepp}{$e^+e^- \to \pi^+\pi^-$}
\newcommand{\ts}{\textstyle}
\newcommand{\ght}[1]{\textcolor{blue}{#1}}
\newcommand{\csk}[1]{\textcolor{red}{#1}}
\newcommand{\glc}[1]{\textcolor{cyan}{#1}}

\bigskip
\vspace{2cm}
\title{Invisible decays of vector Charmonia and Bottomonia\\
to determine the Weak Mixing Angle at quarkonia scale }
\vskip 6ex

\author{G. Hern\'andez-Tom\'e}
\email{gerardo_hernandez@uaeh.edu.mx}
\affiliation{Área Académica de Matemáticas y Física, Universidad Autónoma del Estado de Hidalgo,
Carretera Pachuca-Tulancingo Km. 4.5, Código Postal 42184, Pachuca, Hgo.}

\author{C. S. Kim}
\email{cskim@yonsei.ac.kr}
\affiliation{ {Department of Physics and IPAP, Yonsei University, Seoul 03722, Korea} \\
{CNPR, Dept. of Physics, Chonnam National University, Gwangju 61186, Korea} }

\author{G. López Castro}
\email{gabriel.lopez@cinvestav.mx} 
\affiliation{ {\it Departamento de F\'isica, Centro de Investigaci\'on y de Estudios Avanzados del Instituto Polit\'ecnico Nacional} \\ {\it Apartado Postal 14-740, 07000 Ciudad de M\'exico, M\'exico}}

%\author{D. Sahoo}
%\email{Dibyakrupa.Sahoo@fuw.edu.pl}
%\affiliation{Institute of Theoretical Physics, Faculty of Physics, University of Warsaw, ul. Pasteura 5, %02-093 Warsaw, Poland.}

%\date{6th March 2005}
\bigskip

\bigskip

\begin{abstract}

We compute the branching fractions of vector quarkonia ($V_Q=J/\psi, \psi', \Upsilon(nS)$) decays into neutrino pairs, considering both Dirac and Majorana types, within the Standard Model (SM) and beyond.
The vector nature of quarkonium states yields a decay width in the SM that depends upon the weak vector coupling of the heavy quark, offering the possibility to measure the weak mixing angle at the quarkonia mass scales. If neutrinos have non-standard neutral weak couplings, this could help to distinguish the nature of neutrinos in principle.
 \end{abstract}

\maketitle
\bigskip

\section{Introduction}
Fifty years ago, the discovery of the narrow charmonium state $J/\psi$ \cite{SLAC-SP-017:1974ind,E598:1974sol} pointed to the existence of the charmed quark. Three years later, the $\Upsilon$ bound state of bottom-antibottom quarks was discovered at Fermilab \cite{E288:1977xhf}. Today, a multitude of decay channels of these vector quarkonium states ($V_Q$) and their excitations have been measured (see, for example, \cite{pdg2024}) at flavor factories, primarily at $e^+e^-$ machines that can be tuned to the masses of vector quarkonium resonances, producing large samples of these particles. While quarkonium predominantly decays into hadronic channels, decays into a $\ell^+\ell^-$ pair occur at the few-percent level \cite{pdg2024} and can be measured with higher precision.  In the case of $J/\psi$ meson searches have been reported for a few semileptonic channels. Both in charmonium and botommonium searches of lepton flavor violating decay processes into $\ell^+\ell'^{-}$ ($\ell \neq \ell'$) have also been reported \cite{pdg2024}. 

In this letter, we focus on the calculation of the invisible decay width of $J^{PC}=1^{--}$ quarkonium states, namely of their weak $V_Q \to \nu\bar{\nu}$ decays. Earlier estimates for the invisible decay widths of $\Upsilon$ and $J/\psi$ quarkonia states were reported in Refs. \cite{Chang:1997tq,Bergstrom:1987tg,Ellis:1980ng, McElrath:2005bp, Fayet:2009tv, Yeghiyan:2009xc, Fernandez:2014eja}: {Refs. \cite{Bergstrom:1987tg,Ellis:1980ng} focus on the measurement of the invisible widths within the SM as a tool to derive the number of neutrinos. In Refs. \cite{McElrath:2005bp, Fayet:2009tv, Yeghiyan:2009xc, Fernandez:2014eja}, on the other hand, enhanced missing energy signals in quarkonium decays are aimed to be interpreted as due to dark matter particles. } 
In this work, we specifically investigate the effects of the weak mixing angle (i.e. $\sin^2 \theta_W(\mu)$) on the invisible decay widths  and extend the analysis by considering non-standard neutrino interactions, too. From our study, we suggest that determining the weak mixing angle at the quarkonia scales can be achieved, provided that the invisible widths are measured with relatively good accuracy.
We note that measurements of either the left-right or the forward-backward asymmetries due to $\gamma-Z$ interference in $e^+e^-$ collisions at the quarkonium resonance scale can also provide information on the running of the weak mixing angle. However, such interference effects are expected to be highly suppressed, as the quarkonium mass scale is far from the $Z$-pole.

The invisible width of the $Z$ gauge boson has played an important role in establishing the number of light neutrinos within the SM and setting constraints on invisible decay products beyond the SM. Although the measurement of the invisible width of the $Z$ boson is possible from the difference between the calculated total and visible widths or from distortions in the $Z$ boson lineshape \cite{pdg2024,ALEPH:2005ab}, the invisible width of quarkonium poses a challenge to experiments. 
One of the best ways to measure invisible decay widths would be by using triggering events of visible decays, e.g., at a future Higgs factory:
$H \to Z^* (\to \mu^+ \mu^-) + Z (\to {\rm invisible~~ decay}) $.
Similarly, we can use $\chi_{c0} \to \gamma + J/\psi$ to measure the invisible decay width of $J/\psi$, and $\chi_{b0} \to \gamma + \Upsilon(1S)$
to measure the invisible decay width of $\Upsilon(1S)$.
Searches of $Z'\to {\rm invisible}$ have been reported by Belle-II using the process, $e^+e^- \to \mu^+\mu^-Z' (\to {\rm invisible})$ \cite{Belle-II:2022yaw, Corona:2024xsg}.

Upper limits on invisible decays of some quarkonia states have been reported by the BABAR collaboration. Using a sample of $473\times 10^{6}$ $B\bar{B}$ pairs and the $B\to K^{(*)}J/\psi,\  K^{(*)}\psi(2S)$ the following results were reported: ${\rm BR}(J/\psi\to {\rm invisible})< 3.9\times 10^{-3}$ and ${\rm BR}(\psi(2S)\to {\rm invisible})< 15.5\times 10^{-3}$ at the 90\% C.L. \cite{BaBar:2013npw}. On the other hand, using $91.4 \times 10^6$ $B\bar{B}$ pairs and the $\Upsilon(3S)\to \pi^+\pi^-\Upsilon(1S)$ decay, with the undetectable $\Upsilon(1S)$ recoiling against the di-pion pair, the following result was reported: ${\rm BR}(\Upsilon(1S)\to {\rm invisible})< 3.0\times 10^{-4}$ at 90\% C.L. \cite{BaBar:2009gco}. 
%BES-III has also carried out searches involving invisible particles in decays of $J/\psi$ or vector mesons \cite{BESIII:2018bec, BESIII:2023jji}. 
As this letter will show, the predicted branching fractions of invisible decay widths of vector quarkonia are not completely negligible and could be within the reach of current and future data sets at BES-III and Belle-II experiments. 
{The number of $J/\psi$ and $\psi(2S)$ events collected so far by BES-III Collaboration are, respectively, $(10087\pm 44)\times 10^6$ \cite{BESIII:2021cxx} and $(2712.4\pm 14.3)\times 10^6$ \cite{BESIII:2024lks}. It is also expected that at the Super tau-charm Factory \cite{Lyu:2021tlb}, of the order of $10^{12}$ $J/\psi$ and $10^{11}$ $\psi(3686)$ events could be produced per year. Finally, the Belle-II experiment within its designed luminosity expects to accumulate of the order of $10^{11}$ $\Upsilon(4S)$. This makes possible that the measurements of the invisible ratios of quarkonia decays become feasible in foreseeable future.}
  
\section{Leptonic two-body decays of quarkonia in the SM}
    
\begin{figure}[t]
\vspace{0.0cm}
\begin{tabular}{cc}
\includegraphics[width=7cm, angle=0]{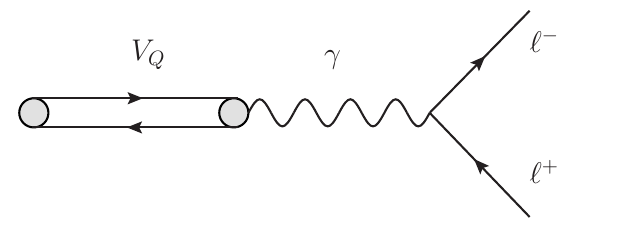}
&\includegraphics[width=7cm, angle=0]{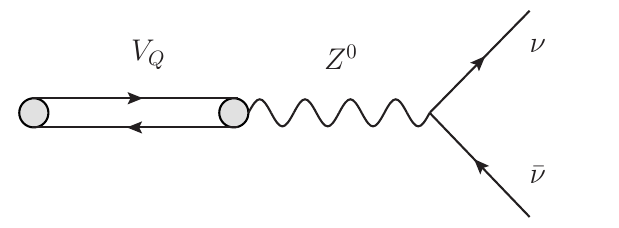}\\
(a)& (b)\\
\end{tabular}
\vspace{.5cm}
\caption{Leptonic decays of heavy $Q\bar{Q}$ vector quarkonium: (a) charged leptons, (b) $\nu\bar{\nu}$ pair. } 
\label{rmo}
\end{figure}

Since we will use it later as a normalization channel, let us first consider the decays of a vector quarkonium into charged leptons $V_Q \to \ell^+\ell^-$ ($V_Q$ is a $c\bar{c}$ or $b\bar{b}$ bound state in a $J^{PC}=1^{--}$ configuration). This decay proceeds in the SM mainly through the one virtual photon annihilation (Figure \ref{rmo}(a)). The coupling of $V_Q$ to the photon is defined as
\be \label{mat-ele}
\langle 0 |j_{\mu}^{\rm em} |V_Q\rangle = q_{_Q} \langle 0 |\overline{Q}\gamma_{\mu} Q|V_Q \rangle = \frac{M^2}{f_{V_Q}} \epsilon_{\mu}\,
\ee
where $j_{\mu}^{\rm em}$ is the electromagnetic current density,  $q_{_Q}$ is the electric charge of the heavy quark $Q$ in units of the proton charge, and $M$ (and $\epsilon_{\mu}$) is the mass (polarization four-vector) of the quarkonium state. Here $f_{V_Q}$ is a dimensionless coupling constant that drives the annihilation probability of the $Q\bar{Q}$ pair by the electromagnetic field. 

The corresponding partial width for decays into charged leptons is well known:
\be \label{rate2l}
\Gamma(V_Q \to \ell^+\ell^-) = \frac{4\pi\alpha^2M}{3f_{V_Q}^2} \left (1+2x_{\ell} \right) \sqrt{1-4 x_{\ell}}
\ee
where $x_{\ell} =(m_{\ell}/M)^2$, and $\alpha$ is the fine structure constant. From this expression and the measured leptonic branching fraction, we can extract the quarkonium annihilation constant $f_{V_Q}$.

\begin{figure}[t]
\vspace{-2.5cm}
\includegraphics[width=16cm, angle=0]{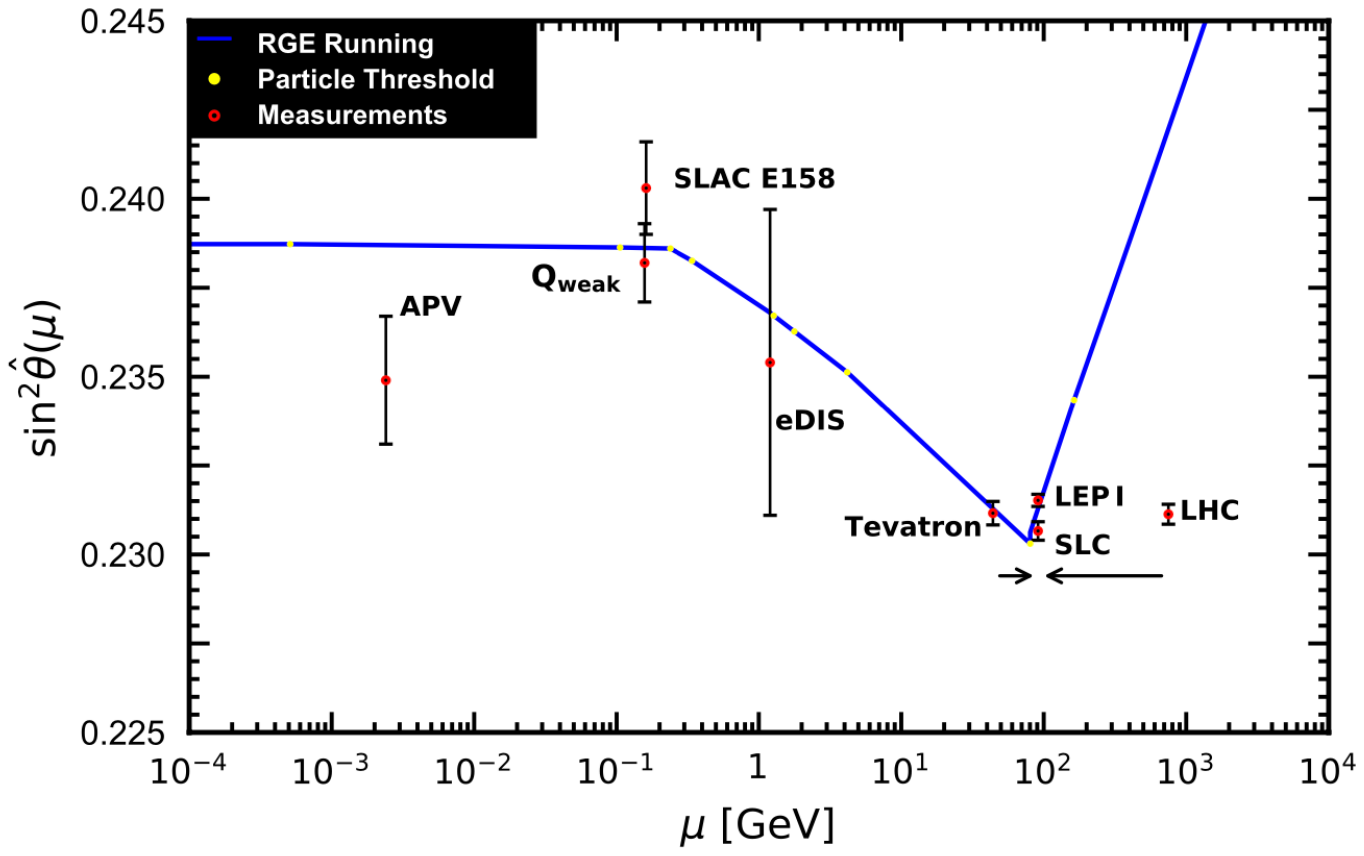}
\vspace{-1.5cm}
\caption{ Scale dependence of the weak mixing angle in the $\overline{\rm MS}$ scheme (taken from Ref. \cite{pdg2024}).}
\label{wman}
\end{figure}

Let us now consider the invisible decays of quarkonia, which are mediated by the neutral weak current (see Figure \ref{rmo}(b)):
\be \label{invisible}
V_Q(q) \to Z^* \to \nu(p)\bar{\nu} (p')\ .
\ee
In the SM, the coupling of the $Z$ boson to fermions is given by
\be
{\cal L}^{NC}=-\frac{g}{\cos \theta_W}J_{\mu}^{Z f}\cdot Z^{\mu}=-\frac{g}{2\cos \theta_W} \left[ \bar{f}\gamma_\mu(g_V^f-g_A^f\gamma_5) f\right]Z^{\mu}\ , \label{zff}
\ee
where, at the tree-level\footnote{The electroweak corrections to the $Zf\bar{f}$ vertex modify these relations \cite{ALEPH:2005ab} to  $\bar{g}_V^f=\sqrt{\hat{\rho}_f}(t_3^f-2q_f \sin^2 \hat{\theta}_W(\mu) )$ and $\bar{g}_A^f=\sqrt{\hat{\rho}_f}t_3^f$, where the effects of electroweak corrections are absorbed into the scale-dependent parameters $\hat{\rho}_f,\, \sin^2 \hat{\theta}(\mu)$.  }, $g_V^f =t_{3}^f-2 q_f\sin^2\theta_W$ and $g_A^{f}=t_3^f$ are the vector and the axial couplings to fermions,  $t_3^f=+1/2~ (-1/2)$ for neutrinos and up-type quarks (for charged leptons and down-type quarks) and $g$ is the SU(2) coupling constant. The parameter $\theta_W$ is the weak mixing angle, and we will use its running values at the relevant quarkonium mass scales in our numerical estimates. Figure \ref{wman}, taken from Ref. \cite{pdg2024}, shows the predicted scale dependence of the weak mixing angle $\sin^2 \hat{\theta}_W (\mu)$ in the $\overline{\rm MS}$ scheme (blue line) \cite{Czarnecki:1998xc, Fanchiotti:1989wv, Erler:2004in} and a few values measured by different experiments. As can be seen, measurements of the weak mixing angle at the quarkonium mass scales have not been reported yet. 

Using the notation in Eq. (\ref{invisible}), the decay amplitude in the case that neutrinos are Dirac particles becomes
%\be \label{Amp-Dirac}
%{\cal M}^D(p,p') = -\left(\frac{g}{2\cos \theta_W} \right)^2 ~ \frac{i}{M^2-m_Z^2} ~ \bar{v}(p') \gamma^{\mu} P_L u(p)  \langle 0|g_V^Q \bar{Q}\gamma_{\mu} Q|V_Q \rangle\ .
%\ee
\be \label{Amp-Dirac}
{\cal M}^D(p,p') = -\left(\frac{g}{2\cos \theta_W} \right)^2  \frac{i}{M^2-m_Z^2} ~ \bar{u}(p) \gamma^{\mu} 
(g_V^\nu-g_A^\nu \gamma_5) v(p') ~ \langle 0|g_V^Q \bar{Q}\gamma_{\mu} Q|V_Q \rangle,
\ee
where in the SM $(g_V^\nu-g_A^\nu \gamma_5)=\frac{1}{2}(1-\gamma_5) \equiv P_L$ is the left-handed projection operator. 
Given the vector nature of quarkonia, only the weak vector current participates in quarkonia annihilation. Here, $g_V^Q$ is the weak vector charge of the heavy quark $Q=b,c$. We keep the finite mass $M$  of the quarkonium state in the $Z$-boson propagator because it is not negligible at least for the bottomonium states. 

The hadronic matrix element in Eq. (\ref{Amp-Dirac}) is clearly related to Eq. (\ref{mat-ele}).  In terms of the  Fermi constant ($G_F/\sqrt{2} =g^2/(8m_Z^2\cos^2 \theta_W))$, we can write the previous amplitude  as follows:
\ba \label{AmpDirFin}
{\cal M}^D (p,p')&=& i\sqrt{2} G_F \left(\frac{g_V^Q}{q_{_Q}}\right) \frac{m_Z^2}{m_Z^2-M^2}~ \bar{u}(p) \gamma^{\mu} (g_V^\nu-g_A^\nu \gamma_5) v(p')~  \frac{M^2}{f_{V_Q}}~ \epsilon_{\mu}(q) \nonumber \\
&=& -i \frac{\widehat{G}_Z}{2}  \epsilon_{\mu}(q)~ \bar{u}(p) \gamma^{\mu}(g_V^\nu - g_A^\nu \gamma^5) v(p') ,
\ea
where $g_V^\nu = g_A^\nu = 1/2$ within the SM,
and we have defined the effective dimensionless constant 
$$\widehat{G}_Z\equiv 2\sqrt{2}~ G_F~ \frac{g_V^Q}{q_{_Q}}~ \frac{M^2}{f_{V_Q}}~ \frac{m_Z^2}{m_Z^2-M^2}~.$$ 
The second line of Eq. (\ref{AmpDirFin}) looks identical to Eq.  (11) in Ref. \cite{Kim:2023iwz} (and in Ref. \cite{Kim:2022xjg}), which is the most general decay amplitude of $Z \to \nu \bar\nu$ with Lorentz invariance, CP and CPT conservation,  by replacing $g_Z $ to $\widehat{G}_Z$. 

If we neglect the tiny masses of neutrinos, we get the following expressions for the invisible width of quarkonium decay in the SM:
\ba 
\Gamma^D_{\textrm{SM}}(V_Q\to \nu\bar{\nu})&=& \frac{\widehat{G}_Z^2 M}{96\pi}.\label{WidthSM}\ 
\ea
Using Eqs. (\ref{rate2l}) and (\ref{WidthSM}) we can estimate the branching fractions of $V_Q\to \nu\bar{\nu}$:
\ba \label{brnn}
{\rm BR}(V_Q\to \nu\bar{\nu})&=& \frac{\Gamma(V_Q\to \nu\bar{\nu})}{\Gamma(V_Q \to \ell^+\ell^-)} \cdot {\rm BR}(V_Q\to \ell^+ \ell^-), \nonumber  \\
&=& N_\nu \left[ \frac{G_F M^2}{4\pi \alpha} \left( \frac{g_V^Q}{q_{_Q}}\right) \frac{m_Z^2}{m_Z^2-M^2}\right]^2 {\rm BR}(V_Q\to e^+ e^-) \ ,
\ea
where in the second equality
$N_\nu$ is the number of active neutrinos, and
${\rm BR}(V_Q\to e^+ e^-)$  denotes the branching fraction for vector quarkonia decays into electron-positron pair for which we use their experimental values (second column in Table \ref{BR-quarkonium}) reported in the PDG \cite{pdg2024}.
Note that the above expression is rather simple and does not depend on the strong interaction effects since the annihilation constant $f_{V_Q}$ has been cancelled in the ratio.
We have used $\alpha=1/137.036$ for the fine structure constant instead of the running $\alpha$ at the relevant quarkonia mass scale, and $N_\nu=3$.

Note that Eq. (\ref{brnn}) has a dependence on the weak mixing angle through $g_V^Q~(Q=b,c)$.
 We have assumed approximated values for the running mixing angles in the $\overline{\rm MS}$ scheme at the relevant quarkonium scales from Refs. \cite{Czarnecki:1998xc, Fanchiotti:1989wv, Erler:2004in}, as shown in the third column.
 Results for the branching fractions of different invisible quarkonium decays within the SM are calculated using Eq. (\ref{brnn}) and are shown in the 4th column of Table \ref{BR-quarkonium}. 
In the last two columns of Table \ref{BR-quarkonium} we have estimated the expected uncertainties for $\sin^2 \hat{\theta}_W(\mu)$ by assuming an  (optimistic) uncertainty of 10\% and 5\%, respectively.
For comparison, the experimentally measured values of BR($Z\to e^+e^-$) and the invisible branching fraction of the $Z$ boson \cite{pdg2024} are also shown in the last row of Table \ref{BR-quarkonium}.
\begin{table}[h] 
 \begin{tabular}{|c|c|c|c|c|c|}
 \hline $Q\bar{Q}$ 
 & BR($V_Q\to e^+e^-$)  & $\sin^2 \hat{\theta}_W(\mu)$ & BR($V_Q \to \nu\bar{\nu}$) & $\Delta \sin^2\hat{\theta}_W$ & $\Delta \sin^2\hat{\theta}_W$  \\
 State & Ref. \cite{pdg2024} & & & $\pm 10\%$  $\Delta B_{\nu\nu}$ & $\pm 5\%$  $\Delta B_{\nu\nu}$ \\
  \hline \hline
 $J/\psi$ & $(5.971\pm 0.032)\times 10^{-2}$ &  $0.237 \pm 0.001$  & $  (2.04 \pm  0.03) \times 10^{-8}$ & $\pm 0.007$ & $\pm 0.003$ \\
 $\psi({\rm 2S})$ & $(7.94\pm 0.22)\times 10^{-3}$ &  $0.236 \pm 0.001$ &  $  (5.52 \pm  0.17) \times 10^{-9}$ & $\pm 0.007$ & $\pm 0.004$ \\
$\Upsilon({\rm 1S})$ & $(2.39\pm 0.08)\times 10^{-2} $ &  $0.233 \pm 0.001$ & $  (1.02 \pm  0.03) \times 10^{-5}$ & $\pm 0.027$ & $\pm 0.015$  \\
$\Upsilon({\rm 4S})$ & $(1.57\pm 0.08)\times 10^{-5}$ &  $0.232  \pm 0.001$ & $  (1.05 \pm  0.05) \times 10^{-8}$ & $\pm 0.029$& $\pm 0.018$ \\
\hline\hline
$Z$ & BR($Z\to e^+e^-$)=  & 0.23129(4) &  BR($Z \to \nu\bar{\nu}$)=    & &  \\
& $(3.3632 \pm 0.0042)\times 10^{-2}$ & &   $(20.000\pm 0.055)\%$  & &\\
 \hline \hline
 \end{tabular}
\caption{Branching fractions of quarkonium decays into a neutrino pair in the SM (4th column), using the reference values of the running mixing angle (third column) in the $\overline{\rm MS}$ scheme at corresponding scales \cite{Czarnecki:1998xc, Fanchiotti:1989wv, Erler:2004in}.  The fifth (sixth) column displays the expected uncertainties in the extracted value of $\sin^2\hat{\theta}_W (\mu)$ by assuming that the invisible branching fraction is measured with 10\% (respectively, 5\%) uncertainty. 
The BR($Z\to e^+e^-$) and invisible ratio of the $Z$ boson are shown for comparison \cite{pdg2024}. 
}\label{BR-quarkonium}
 \end{table}

 A measurement of the invisible width of quarkonium decay would allow a determination of the vector weak coupling $g_V^Q$ (or the running mixing angle $\sin^2 \hat{\theta}_W(\mu)$) at the mass scale of quarkonium.  This is very interesting in itself since no measurements of the weak mixing angle at the quarkonium masses scales are available so far (see Figure 10.2 in the review of the `Electroweak Model and Constraints on New Physics' in Ref. \cite{pdg2024}). As observed in Table  \ref{BR-quarkonium}, charmonium decays would allow a determination of $\sin^2 \hat{\theta} (\mu)$ with similar accuracy to the `eDIS' point in Figure \ref{wman} if their invisible branching fraction is measured with a 5\% accuracy. Invisible decays of $\Upsilon(1S, 4S)$ would require better precision to provide a competitive determination. It is interesting to observe that the branching fraction  of $\Upsilon(1S)\to \ {\rm invisible}$ calculated in Table \ref{BR-quarkonium} is only one order of magnitude below the current upper limit reported by the BABAR collaboration \cite{BaBar:2009gco}. 
 
 {As shown in our Introduction, it is expected to produce the order of $10^{12}$ $J/\psi$ \cite{Lyu:2021tlb}, and $10^{11}$ $\Upsilon(4S)$ \cite{Belle-II:2018jsg} in foreseeable future. By just assuming  1--10\% detection efficiency, one can eventually reach the branching ratios shown in Table \ref{BR-quarkonium} to measure the weak mixing angle at the quarkonia mass scale. We also find the current upper limit on the $\Upsilon(1S)\to $ invisible \cite{BaBar:2009gco} is just an order of magnitude above the predicted rate for $\Upsilon(1S)\to \nu\bar{\nu}$, which is shown in Table \ref{BR-quarkonium}.}
 In the next section we study the interference effects of the weak and electromagnetic contributions at low energies, which could be highly suppressed.

\section{$\gamma-Z$ interference in charged leptonic modes}

From Table \ref{BR-quarkonium} we observe that the invisible width of charmonia are almost six orders of magnitude suppressed respect to their corresponding decays into charged leptons. On the other hand, the invisible widths of bottomonia are only three orders of magnitude smaller compared to their decays into electron-positron. This is mainly due to the quarkonium mass dependence in Eq. (\ref{brnn}). Thus, it is worth exploring in more detail the $\gamma-Z$ interference in $V_Q\to \ell^+\ell^-$ decays and how it would affect the determination of the invisible decay branching fraction ${\rm BR}(V_Q\to \nu\bar{\nu})$,
and henceforth the weak mixing angle $\sin^2\hat{\theta}_W (\mu)$.

The addition of the $Z$ boson contribution to the diagram shown in Figure 1(a), $V_Q \to \ell^+ \ell^-$, modifies Eq. (2) to
%\ref{rmo}$(a)$ we derive the following expression for the leptonic width:
%\be
%\Gamma^{\gamma+Z}(V_Q\to \ell^+\ell^-)=\frac{4\pi\alpha^2M}{3f_{V_Q}^2} \left[ \frac{}{}(V_Q^2+A_Q^2)(1-x_\ell)+3x_\ell(V_Q^2-A_Q^2) \right] \sqrt{1-4x_\ell^2}\ ,
%\ee
\be
\Gamma^{\gamma+Z}(V_Q\to \ell^+\ell^-)=\frac{4\pi\alpha^2M}{3f_{V_Q}^2} \left[ \frac{}{}V_Q^2(1+2x_\ell)+A_Q^2(1-4x_\ell) \right] \sqrt{1-4x_\ell}\ .
\ee
We have defined:
\ba
V_Q&=& 1+ \frac{\sqrt{2}G_F}{4\pi\alpha} \frac{m_Z^2M^2}{M^2-m_Z^2} \left(\frac{g_V^Qg_V^\ell}{q_{_Q}} \right)\, \label{VQ},\\
A_Q&=& \frac{\sqrt{2}G_F}{4\pi\alpha} \frac{m_Z^2M^2}{M^2-m_Z^2} \left(\frac{g_V^Qg_A^\ell}{q_{_Q}} \right)\, ,
\ea
where $g_{V,A}^f$ denote the weak vector/axial coupling of fermion $f$. Clearly, if we turn off the weak interactions, we recover the well known result shown in Eq. (\ref{rate2l}).  Given the quadratic dependence of the leptonic rate upon $V_Q$ and  $A_Q$, only the second term in Eq. (\ref{VQ}) will enter linearly. Since the vector weak coupling of charged leptons is very small ($g_V^{\ell} \approx -0.05$), the $\gamma-Z$ interference would change the decay width very little, by at most $10^{-3}$ in bottomonium decays, and even smaller in charmonium cases, a negligible contribution within the current experimental accuracy.
These small numerical changes would affect the denominator of the first equality in Eq. (8), resulting in almost no impact on our previous numerical analysis presented in Table I.

\section{Comments on invisible decay width in case of  Majorana neutrinos} 

Now we consider quarkonia decays into a neutrino-antineutrino pair in case neutrinos are Majorana particles, but with the SM V-A interactions. After amplitude anti-symmetrization, the decay amplitude becomes:
\bea
{\cal M}^M(p,p') &=& \frac{1}{\sqrt{2}} \left( {\cal M}^D (p,p')-{\cal M}^D (p',p) \right) \nonumber \\
&=& i\frac{\widehat{G}_Z}{2\sqrt{2}} \epsilon_{\mu}(q) ~ \bar{u}(p) \gamma^{\mu} \gamma_5 v(p')   \ .
\eea
The squared amplitudes, averaged over spin states while retaining the finite masses of the neutrinos, are:
\ba
\overline{|{\cal M}^D|}^2&=&\frac{\widehat{G}_Z^2}{6} \left(M^2-m_{\nu}^2\right) \\ 
\overline{|{\cal M}^M|}^2&=& \frac{\widehat{G}_Z^2}{6} (M^2-4m_{\nu}^2)\ .
\ea
As it was noticed in Ref. \cite{Kim:2023iwz}, the difference of probabilities between Dirac and Majorana neutrinos is proportional to the squared mass of neutrinos, in agreement with the Dirac-Majorana confusion theorem \cite{Kayser:1982br}.

To account for possible differences between Dirac and Majorana neutrinos in the presence of non-standard neutrino interactions, we assume the following modified couplings in Eq. (\ref{zff}),
\be \label{nphys}
g_{V,A}^{\nu_{\ell}} \to C_{V,A}^{\nu_\ell} = \frac{1}{2}+\epsilon_{V,A}^{\nu_\ell}.
\ee
Thus, neglecting the masses of neutrinos, Eq. (\ref{WidthSM}) is replaced by:
\ba \label{rate2nu}
\Gamma^D(V_Q\to \nu\bar{\nu})&=& \frac{\widehat{G}_Z^2 M}{48\pi } [(C_V^{\nu_\ell})^2+(C_{A}^{\nu_\ell})^2],\\
 \Gamma^M(V_Q\to \nu\bar{\nu})&=& \frac{\widehat{G}_Z^2M}{24\pi }  (C_A^{\nu_\ell})^2\,. 
\ea
New physics contributions in neutrino couplings (which certainly is very constrained from the invisible width of the $Z$ boson \cite{ALEPH:2005ab}) evade the Dirac-Majorana confusion theorem and lead to a difference of these decay rates\footnote{Note that the SM limit, that is Eq. (\ref{WidthSM}), is recovered for $C_V^{\nu_\ell}=C_A^{\nu_\ell}=1/2.$} proportional to  $\epsilon_V^{\nu_\ell}-\epsilon_A^{\nu_\ell}$.  However, assuming $\epsilon_V^{\nu_\ell}-\epsilon_A^{\nu_\ell} \sim 10^{-1}- 10^{-2}$ gives a difference of branching fractions $2({\rm BR}^D-{\rm BR}^M)/({\rm BR}^D+{\rm BR}^M) \sim 10^{-2}-10^{-4}$, much smaller than the values quoted in Table \ref{BR-quarkonium}, which would require a measurement of the branching fractions 
with such high precision to distinguish between the nature of neutrinos in invisible quarkonium decays
-- that the accuracy needed is out of reach for present facilities worldwide.

%\textcolor{cyan}{(Are reasonable such assumptions in the size of $\epsilon_{V,A}$ in view of the measured $Z\to$ invisible %width???)}

\section{Conclusions}

The peculiar vector feature of quarkonium states makes it suitable for measurements of the weak mixing angle at the mass scales of charmonium and bottomonium states. The branching fractions of their invisible decays are clean predictions of the Standard Model that depend mainly on the weak vector couplings. The invisible decays of $\Upsilon(1S)$ (and $J/\psi$) is predicted to occur at the $10^{-5}$ ($10^{-8}$) level, which seems at the reach of current $e^+ e^-$ colliders, like Belle-II (BES-III). A meaningful determination of the weak mixing angle at those scales would require measurements with a few percent accuracy 

Invisible widths of quarkonia can also be useful to investigate the nature of neutrinos. In the presence of non-standard neutrino interactions that modify the vector and axial couplings of the $Z$ boson to neutrinos in a different way, it appears a difference between the decay probabilities of Dirac and Majorana neutrinos that evades the Dirac-Majorana Confusion theorem \cite{Kayser:1982br}. Distinguishing Dirac and Majorana neutrinos seems promising again in decays of $\Upsilon(1S)$ and $J/\psi$ if there exists new physics beyond the SM, but it also requires measuring its invisible width with much exquisite precision. 

Our work focuses on the determination of the weak mixing angle at the quarkonia scales, which so far are missing. Invisible decays for this purpose have not been considered in any literature before, and are of great importance in order to establish the evolution of the weak mixing angle  at lower energy scales. Precise measurements of decay fractions would indeed provide a competitive determination at those scales, but even a not-so-precise measurement could give us for the first time indications of the central value of the weak mixing angle at those scales.

As a final comment, we note that neutrino scattering off nuclei can also provide an information about the weak mixing angle at low energies. The NuTeV collaboration has achieved an accurate determination of $\sin^2 \theta_W$ using deep inelastic scattering of neutrinos and antineutrinos on isoscalar nuclei at an average scale of $\langle Q \rangle \approx 5$ GeV \cite{NuTeV:2001whx}. Their result extrapolated to the $Z$ boson mass scale $\sin^2 \theta_W(m_Z^2)\big|_ {\overline{\rm MS}}=0.2356(16)$ \cite{NuTeV:2001whx,Kumar:2013yoa} is 3$\sigma$'s above the value expected in the SM. 
Eventually, coherent neutrino scatterings on nuclei will provide another method to determine the weak mixing angle at very low energies \cite{COHERENT:2017ipa}, with a precision that is limited by the difficulty measuring the very small nuclear recoil.  The determination using the invisible width of quarkonia states, which is discussed in this paper, would provide results for the weak mixing angle at the mass scale of those quarkonia states, with a precision limited only by the uncertainties in the branching fractions. Together with NuTeV and other low energy measurements \cite{Kumar:2013yoa} as well as our proposed method, a test of the running weak mixing angle at the few GeV scales can be finally achieved.

\section{acknowledgments }
This work of C.S.K. was supported by grants from the National Research Foundation (NRF) of the Korean government 
(RS-2022-NR074767), (RS-2022-NR070836).
G.L.C. acknowledges financial support from Conahcyt project CBF2023-2024-3226.

\end{document}